\documentclass[a4paper,10pt]{article}

\usepackage[utf8x]{inputenc}
\usepackage{amsmath} 
\usepackage{amsfonts}
\usepackage{graphicx}

\title{Gravity and Complexity}
\author{Yves Gaspar\\
Universit\`{a} Cattolica del Sacro Cuore\\
(Catholic University of the Sacred Heart)\\
Department of Mathematics and Physics, \\
Via Musei 41,\\
Brescia, Italy\\
\and Giovanni Acquaviva\\
Universit\`{a} degli Studi di Trento\\
Department of Physics, \\
Via Sommarive 14\\
Trento, Italy}
\date{}

\begin{document}
\maketitle

\abstract{We present a heuristic analysis of the dynamics of general
 solutions to the Einstein Field Equations which highlights the
 possibility that such systems could possess a degree of unpredictability stronger
 than that which characterises chaotic systems. Questions regarding
 features of the complex dynamics of such cosmological models can be
 {\it{undecidable}}. These systems could be qualitatively compared with Turing machines in the sense that even if initial conditions for a dynamical system
 associated to general solutions to the Einstein Field Equations were
 known exactly, then the subsequent evolution could still be
 unpredictable.  }

\section{Introduction.}

Chaotic behaviour occurs in a wide variety of physical systems. In
 Newtonian gravity the three body problem is known to exhibit chaos. 
Spin orbit couplings between particles, accelerator beams and other similar systems can also display chaotic behaviour. In the context of Einstein's General theory of Gravitation the so-called
 Bianchi type $VIII$ and $IX$ models without scalar fields can be chaotic \cite{G},\cite{B},\cite{Hor} and \cite{ring}. The first pioneering work on these general solutions to the Einstein Field Equations has been carried out by Belinski, Khalatnikov and Lifshitz (BKL) \cite{BKLI}, \cite{BKLII} and \cite{BKLIII}. Despite the approximate nature of the analysis of the Eintein Field Equations in work of BKL, it seems that several aspects are strongly supported by various rigorous and numerical approaches. The analysis we present in this paper is heuristic and we argue that our approach can guide more rigorous investigations.\
One might suspect that these anisotropic and homogeneous models are physically irrelevant since our universe is well described on large-scales by isotropic and homogeneous Friedmann-Robertson-Walker (FRW) models. However, a cosmological theory ought to explain why our universe is close to an FRW model today, because the FRW models correspond to a set of measure zero in the space of all possible physically admissible solutions to the Einstein Field Equations (EFEs): in this sense the FRW solutions are extraordinarily special and one ought to understand why our universe is described by the most ``improbable" and symmetrical models. One of the possible scenarios which tries to explain this corresponds to the very important inflationary models. These models claim that, even starting initially with a non-symmetrical and chaotic initial state, through inflation the universe will at late-time approach an FRW solution. However, there is no rigorous proof of this mechanism and in the most general chaotic and 
inhomogeneous case, it is even possible that inflation will not start (chaotic and eternal inflationary solutions are models which try to tackle this problem). The general Bianchi models offer a leading-order approximation to the most general inhomogeneous solutions to EFEs and in order to grasp fully the dynamics which leads to an FRW model these solutions ought to be considered. Furthermore, even if our universe today is described on large scales by an FRW model, in the past very close to the initial singularity space-time can behave like in a chaotic BKL model, typical of Bianchi type VIII and IX, but also typical of general inhomogeneous models. Moreover, realistic singularities inside black holes are of BKL-type, even if from an outside point of view the solution is highly symmetrical. Furthermore, from a rigorous point of view, it can be shown that a Bianchi type VIII model can be close to an FRW model for an arbitrarily long interval of  cosmic clock time: locally, the solution could look like an 
almost flat FRW model. Also, many general Bianchi models, including type VIII, can be viewed as the result of a superposition of gravitational waves on a simpler more symmetrical background space-time and as such they can provide very interesting insights into the physics of space-times containing strong gravitational waves, as might have occurred close to the initial singularity.    
These are some of the interesting arguments that show why solutions more general than FRW models ought to be considered as part of physical models. Furthermore, by excluding general models, one might miss an important ``message" that might be hidden deeply within gravity theory and the EFEs and therefore the possibility of G\"{o}del undecidability within the physical equations for gravity ought to be considered. It is of some importance to note that profound limits to computability can exist in the context of our modern gravitational theory. \\

  The equations of motion for the chaotic systems that we will consider can not be integrated and no
 general formula exists which gives their state at all times. Physicists
 study various approximations and simulations of these systems and various
 statistical or scaling properties, escape rates and other average
 quantities can be computed, because the trajectories characterising the
 evolution of these systems are essentially random. However, as is studied
 by C. Moore in \cite{M}, dynamical systems exist in which these
 computations of average quantities are impossible: their dynamics is not only
 random, it exhibits an additional degree of {\it complexity}. In
\cite{M} an example is constructed of a three-dimensional potential in
 which the motion of a single particle possesses this ``new" degree of
 complexity which is related to the {\it{undecidability}} regarding some
 features of the system. For such complex systems the basins of
 attraction are not merely recursive. The {\it{undecidability}} regarding
 features of the dynamics causes these sets to be more complex than fractal
 sets: at every scale of magnification, qualitatively new behaviour shows
 up. Even if the initial conditions were known exactly, the evolution of
 the system would still be unpredictable because of this additional
 degree of complexity which superposes itself onto the chaoticity of the
 system.\\
In \cite{M} it is argued that such systems can be understood by
 comparing them with Turing machines. These machines correspond to idealised
 computers consisting essentially in a box containing a finite number of
 states and an infinite ``tape" on which sequences of symbols may be
 written. The machine can read only one symbol for a given position on the
 tape and as a consequence it can alter its internal state, change the
 tape symbol and move one space left or right on the tape. Turing machines
 are capable of universal computation: for any given program, there
 exists a Turing machine which will perform it using the tape as its
 registers and memory, such that it will be capable of any finite computation.
 An interesting problem regarding Turing machines is the so-called
 {\it{halting problem}}: given some initial state, will the machine ever halt? For instance, one could construct a Turing machine which searches
 for counterexamples to Riemann's hypothesis and such that it would halt only if it finds one.            
Alan Turing proved this kind of question to be undecidable: in order to
 answer this halting problem one would have to prove (or disprove)
 Riemann's hypothesis. Thus one needs more information than is actually
 available to solve the problem. Consider the set $H$ of sequences on which the
 Turing machine will eventually halt: Rice \cite{R} proved that
 virtually any question about $H$ is undecidable, such as whether $H$ is finite,
 dense, non-empty etc. Even a measure of the set $H$ can not be
 computed. From a dynamical systems viewpoint, these are questions regarding a
 basin of attraction whose features can be undecidable. Furthermore,
 this implies that Turing machines are unpredictable even if the initial
 conditions are known exactly. \\
In this work we will first overview the particular nature of the chaotic
 dynamics of the homogeneous Bianchi type $VIII$ cosmological models.
 These models have been shown to exhibit an interesting type of {\it{time
 asymmetry}} \cite{GT}: their behaviour towards the past is chaotic,
 whereas their behaviour towards the future, away from the initial
 singularity, is characterised by non-chaotic oscillations. This
 behaviour will be reconsidered from a new perspective using the orthonormal frame approach and the Hamiltonian formulation of the problem and it will be shown how these particular properties can lead to new features of the dynamical system related to
 undecidabilty. \\

\section{Hamiltonian evolution in minisuperspace.}

It turns out to be useful to discuss the problem of the time evolution
 of type $VIII$ models in the so-called {\it{Hamiltonian formalism}}
 \cite{M}, \cite{WE}. 
The features of the type $VIII$ model presented in this section can
 also be derived from the results which were obtained by H. Ringstr\"{o}m and
  J.T. Horwood, J. Wainwright et al. \cite{Hor} and \cite{ring} using a rigorous analysis
 of the system, but we 
discuss these features differently in order to stress particular
 characteristics of the dynamics in a clear and transparent way: in the
 subsequent section these properties will be used to derive new
 results, which can not be found elsewhere in the literature.\\

In the Hamiltonian picture the evolution of the type $VIII$ model
 corresponds to the motion of a point particle in a two dimensional billiard,
 where the billiard walls correspond to a triangular shaped potential
 in minisuperspace 
$({{\beta}^{+}}, {{\beta}^{-}})$ \cite{WE}. The parameters   ${{\beta}^{+}}$
  and ${{\beta}^{-}}$ are related to the metric in the following way:
 considering the metric components to be the basic variables for the
 gravitational field, the general form for a line element for Bianchi class A
 models can be written as
$${{ds}^2}= -N(t^{'})d{{{t^{'}}^2}}+{g_{ab}}{W^{a}}{W^{a}}$$
where $ {W^{a}}$ are time-independent one-forms dual to suitable frame
 vectors $e_{a}$. Three time-dependent scale factors can be introduced
 such that 
$$g_{ab}=diag(a^2, b^2, c^2)$$
which can be rewritten as 
$$g_{ab}=diag(e^{2{{\beta}_{1}}},e^{2{{\beta}_{2}}} ,
 e^{2{{\beta}_{3}}})$$
with
$$ {{\beta}_{1}}={{\beta}^{0}} -2{{\beta}^{+}}$$
$$ {{\beta}_{2}}={{\beta}^{0}}+ {{\beta}^{+}}+{\sqrt{3}}{{\beta}^{-}}$$
           
$$ {{\beta}_{3}}={{\beta}^{0}}+ {{\beta}^{+}}-{\sqrt{3}}{{\beta}^{-}}$$

The triangular shaped potential in minisuperspace in which the universe
 point will evolve has also an infinite open channel along the
 ${{\beta}^{+}}$ axis. The triangular shape of the potential allows 
the evolution to be {\it{chaotic}}: the universe point can bounce off
 the walls in a chaotic sequence, in principle {\it ad infinitum}, such that
 the so-called mixmaster behaviour takes place. Now if the type $VIII$
 system is studied towards the past, close to the initial singularity, 
then the triangular potential expands and it has been shown that an
 infinity of bounces are possible which lead to a chaotic evolution. If one
 studies the evolution towards the future, far from the initial
 singularity, then the triangular potential contracts and one would expect that
 the universe point could bounce off the walls in an infinite chaotic
 sequence as well. 
However, our previous analysis  \cite{G} based on a
 combination of the Hamiltonian formalism and the orthonormal frame approach
 shows that this can not happen: unlike in a ``classical Newtonian"
 billiard, the universe point is forced to leave the triangular region of the
 potential and to escape along the infinite open channel along the
 ${{\beta}^{+}}$ axis, 
such that ${{\beta}^{-}} {\rightarrow}   0$ and  ${{\beta}^{+}}
 {\rightarrow}  +{\infty}$. This late-time evolution is characterised by
 {\it{non-chaotic}} oscillations along the two walls of the infinite channel. Note that
 the channel becomes increasingly narrow as ${{\beta}^{+}} {\rightarrow} +{\infty}$
 and the universe point will exhibit increasingly rapid non-chaotic
 oscillations about the ${{\beta}^{+}}$ axis. 
As a consequence the so-called shear parameter ${{\Sigma}^2}$, given by
$${{\Sigma}^2}={{{\sigma}^2}\over{3H^2}}$$
with ${\sigma}^2$ being the shear scalar and $H$ being the Hubble
 parameter, exhibits increasingly rapid non-chaotic oscillations. In fact, if one defines

$${{{\Sigma}_{\pm}}}={{{{\sigma}_{\pm}}}\over{H}}\ ,$$ with 
$${{\sigma}_{+}}={{1}\over{2}}({{\sigma}_{22}}+{{\sigma}_{33}})$$
$${{\sigma}_{-}}={{1}\over{2{\sqrt{3}}}}({{\sigma}_{22}}-{{\sigma}_{33}})\ ,$$

then one can show that, as ${\tau} {\rightarrow} +{\infty}$ (note that the expression is slightly different from the one used in \cite{G},\cite{B},\cite{Hor} and \cite{ring}),
\begin{equation}
\label{six}
{{\Sigma}_{-}}(\tau) {\sim} {e^{-f(\tau)}}osc(b{e^{a{\tau}}} +{\phi})
\end{equation}
where $osc(\tau)$ represents some bounded non-chaotic oscillatory function, $a$ and
 $b$ are suitable constants and ${\phi}$ represents a phase term (as we will explain below, the present analysis will focus on this phase term) and where ${e^{-f(\tau)}}$ can not decrease more quickly than any
 exponential function. The exact asymptotic form for ${{\Sigma}_{-}}(\tau)$ can be found in the work of J.T. Horwood, J. Wainwright et al. \cite{Hor} but for our purpose only the qualitative expression given by equation (\ref{six}) will be sufficient because we will examine the motion of the universe point only as it enters the open infinite channel. \\

Let us note that unbounded scalars can be derived from the {\it{dimensionless
electric and magnetic parts of the Weyl curvature}} \cite{WE}, 
$${{\tilde{E}}_{ab}}={E_{ab}}{{1}\over{H^2}}$$
$${{\tilde{H}}_{ab}}={H_{ab}}{{1}\over{H^2}}$$

For all Class A models one can define 

$${{\tilde{H}}_{+}}={{1}\over{2}}({{\tilde{H}}_{22}}+{{\tilde{H}}_{33}})$$
$${{\tilde{H}}_{-}}={{1}\over{2{\sqrt{3}}}}({{\tilde{H}}_{22}}-{{\tilde{H}}_{33}})$$

and likewise for ${{\tilde{E}}_{\pm}}$.

The magnetic part of the Weyl curvature tensor describes {\it{intrinsic
 general relativistic effects}} which have {\it{no Newtonian
 counterpart}}: the study of its properties is thus important in order understand
 profound differences between Newton's and Einstein's theory of gravity.\\

Using the relations between the expansion normalised
 variables and the Hamiltonian variables, one can show that there exists an
 $n$th order derivative of  ${{\tilde{H}}_{\pm}}$ which will diverge as
 ${\tau} {\rightarrow} +{\infty}$. 
As a consequence, one can show that small phase differences lead to 

$$|{{{H}_{\pm}}^{(n)}}({\phi}_{1})-{{{H}_{\pm}}^{(n)}}({\phi}_{2})|
 {\rightarrow} +{\infty}$$

if $|{{\phi}_{1}}-{{\phi}_{2}}| < {\epsilon}$ for any ${\epsilon} > 0$.
It has also been shown by J. Wainwright \cite{W} that the expansion
 normalised Weyl parameter $W=
 {{{\tilde{H}}_{+}}^2}+{{{\tilde{H}}_{-}}^2}+{{{\tilde{E}}_{+}}^2}+{{{\tilde{E}}_{-}}^2}$ diverges as ${\tau}
 {\rightarrow} +{\infty}$, and one can show that               
 slightly different initial conditions can lead to important
 differences in the evolution of the Weyl curvature parameter.\\
In what follows, we will use the general properties presented in this
 section to show how {\it{undecidability}} emerges in the dynamics of the
 type $VIII$ system. \\

\section{The boundary between chaos and order.}

The analysis of the dynamics of the type $VIII$ cosmological model
 can lead to undecidability. Recall that a formal system is decidable if, for every statement $S$, one can prove whether it is true or false.\\
 Consider a set of initial conditions which in the Hamiltonian picture would correspond to points lying in a region at the beginning of the open infinite channel of the type $VIII$ potential in minisuperspace.
 
Consider a point $p_{i}$ in such a region $R_{1}$
 close to the channel: this point will lead to non-chaotic oscillations between the
 two walls of the channel with an associated phase ${\phi}_{i}$, see the
 discussion in the previous section. 
Around the point $p_{i}$, there exists a neighbourhood (of non-zero measure) of points $q_{i}$ with phase factors
 ${{\phi}^{\prime}}_{i}$ such that $ |{{\phi}_{i}}-{{{\phi}^{\prime}}_{i}}|  <
 {\epsilon}$, for any given $ {\epsilon}$: 
this means that points belonging to this neighbourhood in $R_{1}$ will
 not display sensitivity to initial conditions and the oscillations will display only small phase differences (see points in blue of figure 1).
\begin{figure}[h!]
 \begin{center}
  \includegraphics[width=230pt]{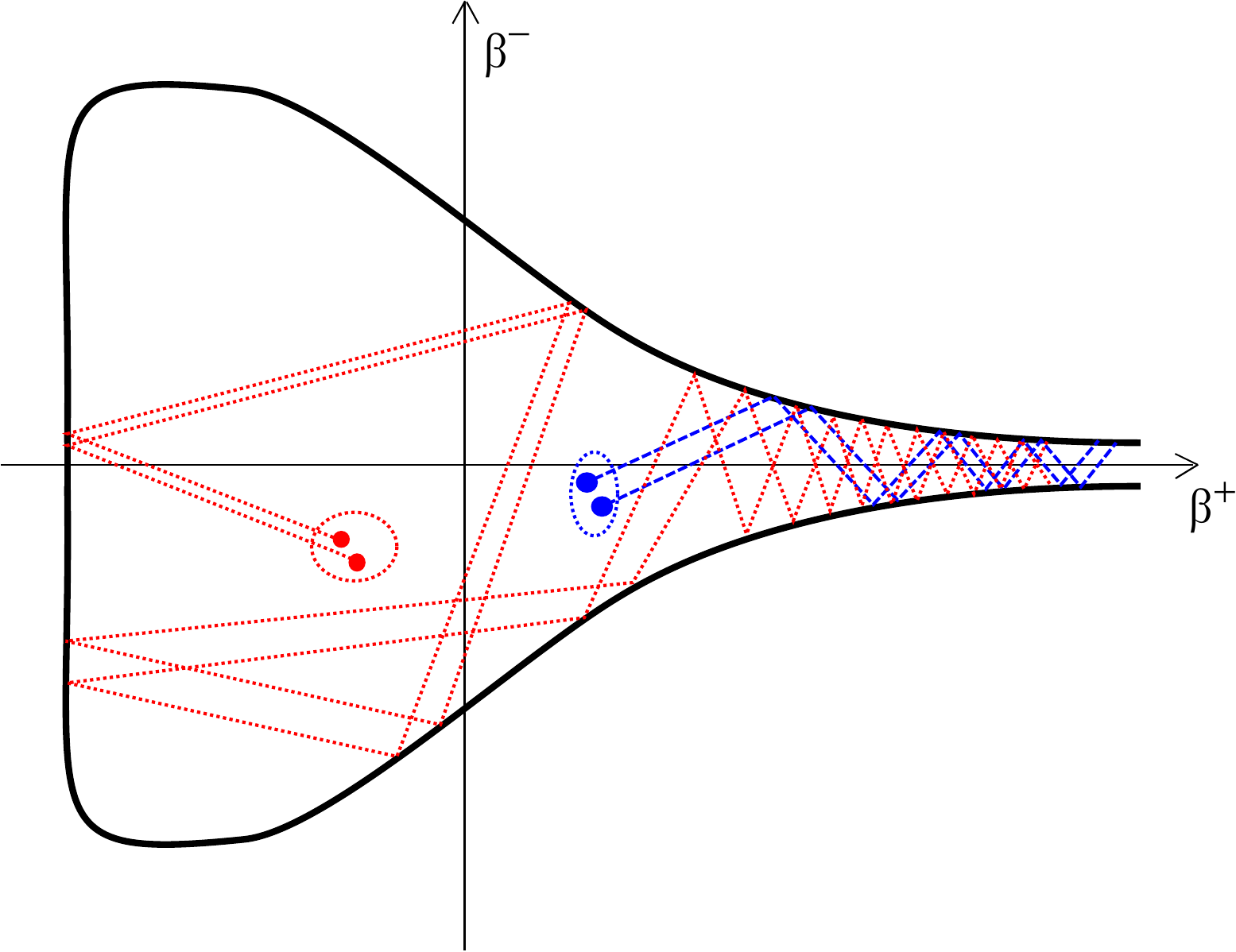}
  \caption{}
 \end{center}
\end{figure}
 
However if one considers the region outside the
 channel in the triangular part of the potential, then
there exist points  $p_{i}$ in a region $R_{2}$ such
 that initial conditions belonging to a neighbourhood (of non-zero measure) of $p_{i}$ will
 lead to chaotic trajectories characterised by a strong sensitivity to
 initial conditions and the resulting phase differences between
 oscillations when the universe point enters the open channel 
would be essentially random and unpredictable (see the points in red in figure 1). One could construct a
 basin of attraction by choosing some suitable value of the phase term
 ${\phi}_{0}$ and consider initial conditions which lead to a phase term
  ${\phi}_{i} < {\phi}_{0}$ or ${\phi}_{i} > {\phi}_{0}$.
For initial conditions belonging to the non-chaotic region $R_{1}$ one
 would not obtain a fractal structure, while for points belonging to the
 chaotic $R_{2}$ region and which would be sufficiently close to the initial
 singularity, the basin of attraction 
would approach a fractal set.  \\
Thus within the same dynamical system, part of the set of initial
 conditions leads to non-chaotic behaviour with respect the phase differences, while the other part is
 characterised by chaotic behaviour with respect to the phase factor ${\phi}$. In fact, as is represented in figure 2, the evolution backwards in time towards the initial singularity of two neighbouring points belonging to a region $R_{1}$ will lead to two points which are no longer close to each other in minisuperspace: the blue points will merge with red points. The closer one approaches the initial singularity, the stronger the mixing will be between points originating within $R_{1}$ and $R_{2}$ regions.\\
\begin{figure}[h!]
 \begin{center}
  \includegraphics[width=230pt]{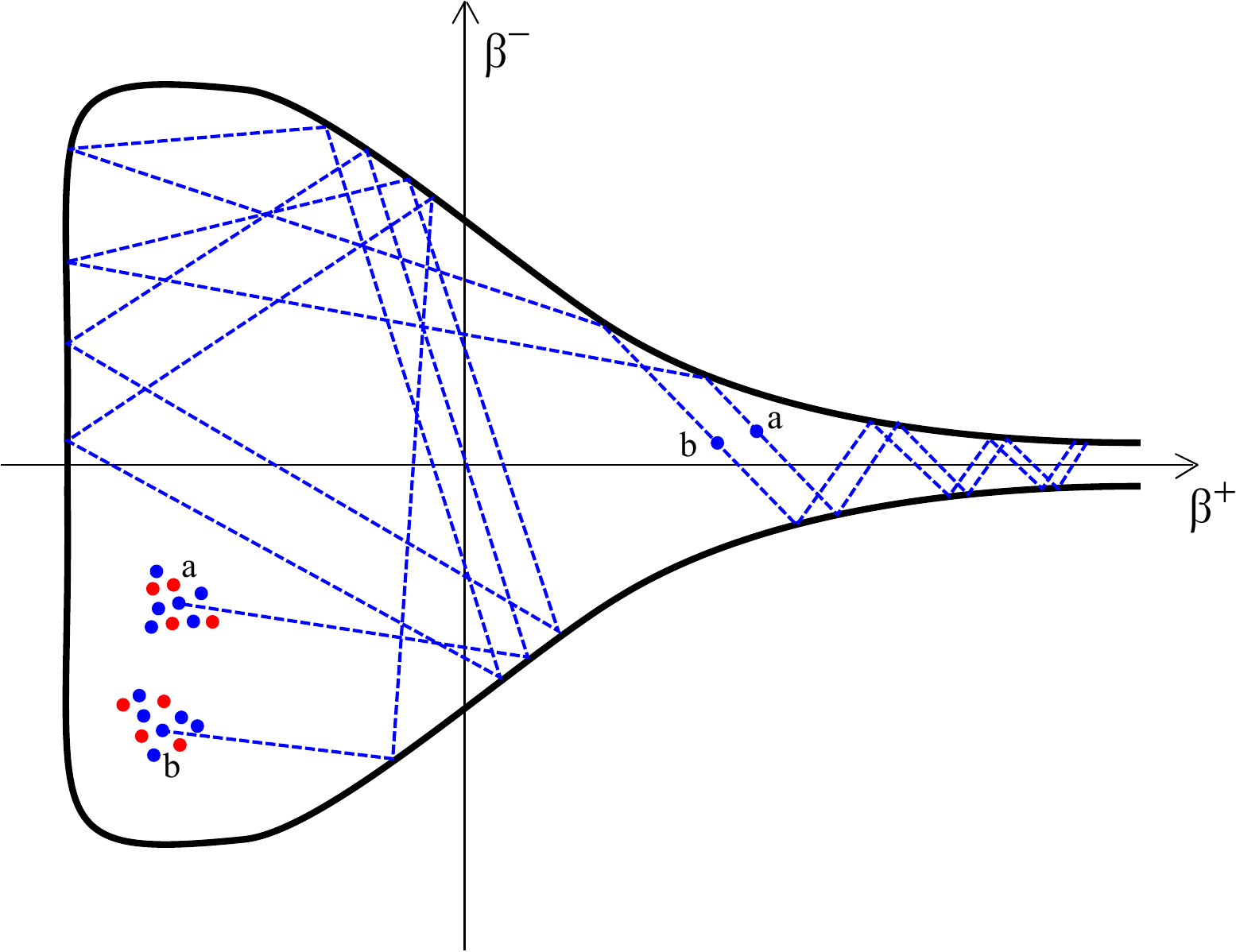}
 \caption{}
 \end{center} 
\end{figure}

This implies that it would be hard to determine whether
 given initial conditions close to the initial singularity lead to chaotic behaviour with respect to the
 phase or not. Chaoticity itself {\it{would depend on the initial
conditions}} in a highly complex way. Therefore, if one studies the type $VIII$ system, using
 some approximation or map (the type $VIII$ equations are non-integrable in the general case of interest), {\it{even if initial conditions would be known
 exactly}}, then the subsequent evolution could still be unpredictable.
 For initial conditions close to the initial 
singularity, the chaoticity itself would be {\it{undecidable}}: for any
 map, approximation or numerical simulation to be meaningfull when trying to tell something
 about the late-time behaviour, even knowing initial conditions exactly,
 one would have to know whether or not the late-time behaviour is
 chaotic. 
But this is what one tries to find out using this approximation of the system.  Thus
 one needs more information then is actually available to solve the
 problem: in order to determine the future behaviour, one would have to
 know not only initial conditions 
as accurately as possible, but also the dependence of the escape rates
 on all the initial conditions should be known and this would require
 knowledge of the exact full solution to the Einstein Field Equations for the
 type $VIII$ model, which is impossible. \\

For initial conditions close to the initial singularity, the future behaviour of the unbounded scalars such as the Weyl parameter presented in the previous section would be unpredictable in the same sense as explained above, even if these initial conditions were known exactly, because these scalars at late times are function of a phase factor ${\phi}$ which depends in an undecidable way on the considered initial conditions. Questions regarding the future behaviour of the dynamics are analoguous to questions about the future behaviour of Turing machines: in a sense the space-time in a neighbourhood of the singularity can be compared with a universal Turing machine, as far as phase differences between late-time oscillations are concerned. \\
Also, it would have no meaning to try compute any average statistical
 quantity over all initial conditions, since the full dynamics would not
 just be random: the basins of attraction would not be purely recursive
 fractal sets,  
an additional degree of complexity would superpose itself on the
 chaoticity of the dynamics. 
The nature of these fractal basins might correspond to the recently
 studied notion of superfractals \cite{SF}. Within such sets the properties
 of the fractals possess an additional random variability. \\
In the next section we will try to compare the degree of complexity or
 unpredictability in Newton's and in Einstein's theory of gravitation.
 \\
 
\section{ Comparing the complexity of Newton's and Einstein's theory of Gravity.}

Surprisingly, an infinity of oscillations in a finite time interval can also occur in the Newtonian many body problem, see the work of J. Xia \cite{JX}. One might then suspect that complexity or undecidabilty could occur as well in Newtonian gravity, because this type of oscillations play a crucial role in the dynamics of general homogeneous solutions to the Einstein Field Equations. However, the configuration which would lead to an infinity of oscillations in a finite interval of time is special: one has to consider four particles of equal mass forming two binary pairs with opposite angular momentum and orbiting on two different parallel planes. Next a fifth lighter particle has to be considered oscillating between the centres of the two binary pairs along a line perpendicular to the two planes: for this configuration it has been shown that the lighter particle will undergo an infinity of oscillations in a finite time. This means that, since the initial configurations leading to an infinity of oscillations 
are special, knowing the initial conditions {\it{exactly}} one could predict the future evolution completely: the Newtonian gravitational system can thus not be compared with the type $VIII$ dynamical system or with Turing machines. Furthermore, the complexity that emerges in the dynamics of solutions to the Einstein Field Equations which we discussed in the previous sections is related to the {\it{time asymmetry}} exhibited by those solutions: the Bianchi type $VIII$ model is chaotic towards the past and it can be shown to be non-chaotic towards the future. This duality regarding chaotic behaviour is linked with the complexity of the dynamics of those models. One would not expect such a level of complexity to emerge in Newtonian many body problems because the solutions exhibit {\it{time symmetry}} between past and future.\\
Another reason which might imply that Newtonian gravity does not lead to undecidabilty regarding features of the dynamics is the following: as explained in section $2$, although the Weyl curvature parameter and other scalars derived from it are known to be unbounded towards the future, the dependence of the phase term associated to their late-time oscillations on the initial conditions can be undecidable. The magnetic part of the Weyl curvature describes intrinsically general relativistic effects which have {\it{no Newtonian counterpart}}: one might thus suspect that the dynamical features exhibited by those scalars possess no Newtonian analogue. \\

\section{Conclusion}

Since the work of K. G\"{o}del \cite{Go} on incompleteness and
 undecidability, several important examples of undecidable propositions in pure mathematics
 have been found. But also in the context of systems which might be
 relevant for physics \cite{JDB} interesting results have been obtained: an example is provided by the work of F. Doria and N. da Costa \cite{DC}. In the latter it was shown that it is impossible {\it{in general}} to demonstrate the stability or instability of equilibrium points of differential equations: the stability is in general undecidable. In order for these results to be of physical relevance, the equilibria have to involve the interplay of a very large number of different forces: however, this situation has not arisen yet in real physical problems. In the present work, our heuristic analysis has shown that undecidability can arise in a physically important problem (although it regards only some details of the dynamics): the study of general initial conditions in cosmology. For a Bianchi type $VIII$ model, the question whether given sets of initial conditions close to the initial singularity lead to chaotic behaviour with respect to phase differences or not can be undecidable: unbounded scalars such 
as the Weyl parameter at late times are function of a phase term which depends in an undecidable way on the initial conditions. Even knowing the initial conditions exactly, one would not be able to predict the future dynamics fully because uncertainties in the phase differences can lead to a different evolution of the unbounded scalars associated to the type $VIII$ solution. The fact that one can not solve the equations for the type $VIII$ model {\it{exactly}} implies not only (usually inevitable) quantitative errors in the study of the future evolution, but even important qualitative errors will emerge: the error can correspond to exchanging chaotic and non-chaotic behaviour. \\
It would be interesting to try to apply the techniques used in \cite{DC} to show rigorously that stability with respect to phase factors is undecidable for the type $VIII$ differential equations, and hence that chaos is undecidable as well. Also, numerical simulations might be able to show the presence of non-recursive fractal sets in the type $VIII$ dynamics. \\
The type $VIII$ solution might provide a leading-order approximation to part of the general inhomogeneous solution to the Einstein Field Equations and one might suspect that a similar level of complexity could occur as well for inhomogeneous solutions. Note that the presence of the singularity is crucial in determining the undecidability with regard to the features of the type $VIII$ dynamics discussed in section 3. One might then argue at a heuristic level, that by eliminating the singularity by an appropriate high-energy physics theory, one could also erase these fundamental problems and obtain a consistent and decidable cosmological model at all levels. However, theories such as quantum gravity or some other high-energy physics theory make use of quantum physics (quantum field theory), which introduces a fundamental unpredictability and could even be itself undecidable, see \cite{Ko} where undecidability in quantum field theory was discussed, see also \cite{GH} where it is shown that calculations of a 
wave-function for a cosmological quantity can turn out to be uncomputable.\\

\section{Acknowledgements.}

I thank Prof. J. D. Barrow for pointing out useful arguments and references and I am grateful to the Trinity Hall Computing Service (Cambridge University, UK) which allowed me to use the computing facilities.


\begin{thebibliography}{99}

\bibitem{G}  J. D. Barrow and Y. Gaspar, {\it{Bianchi type $VIII$
 Empty Futures}}, Class. Quantum Grav. 18, 1809-1822 (2001)\\
\bibitem{B} J. D. Barrow, Phys. Reports 85, 1 (1982)\\
\bibitem{Hor} J.T. Horwood, M. J. Hancock, D. The and J. Wainwright,
 {\it{Late-time asymptotic dynamics of Bianchi $VIII$ cosmologies}},
 gr-qc/0210031 (2002)\\
\bibitem{ring} H. Ringstr\"{o}m, {\it{The future asymptotics of Bianchi
 $VIII$ vacuum solutions}}, gr-qc/0103107 (2001)
\bibitem{BKLI} V. A. Belinski, I. M. Khalatnikov and E. M. Lifshitz, Adv. Phys. 19, 525 (1970)\\
\bibitem{BKLII} V. A. Belinski, I. M. Khalatnikov and E. M. Lifshitz, Sov. Phys. JETP lett., 11, 123 (1970)\\
\bibitem{BKLIII} V. A. Belinski, I. M. Khalatnikov and E. M. Lifshitz, Sov. Phys. USPEKHI, Vol. 13, 6, 745 (1971)\\
             
\bibitem{M}  C. Moore, {\it{Unpredictability and Undecidability in
 Dynamical Systems}}, Phys. Rev. Lett. 64, n. 20, 2354-2357 (1990) 

\bibitem{R} H. Rogers, {\it{Theory of recursive functions and
 effective computability}}, Mc Graw-Hill, New York (1967)

\bibitem{GT} Y. Gaspar {\it{Time Asymmetry and Chaos in General
 Relativity}}, Gen. Rel. Grav. 36, 2085-2094 (2004)

\bibitem{WE} J. Wainwright and G. F. R. Ellis (eds.), {\it{Dynamical
 Systems in Cosmology}}, Cambridge University Press, (1997)

\bibitem{W} J. Wainwright, Gen. Rel. Grav. 32, 1041 (2000)

\bibitem{NJ} N. Cornish and J. Levin, {\it{The mixmaster universe is unambiguously chaotic}}, Proceedings of the 8th Marcel Grossmann meetings, Jerusalem, June 1997, gr-qc/9709037\\
N. Cornish and J. Levin, {\it{The mixmaster universe: a chaotic farey tree}}, Phys. Rev. D55, 7489-7510 (1997) 

\bibitem{JDB} J. D. Barrow, {\it{G\"{o}del and Physics}}, Paper
 delivered at 'Horizon of Truth' for the K. G\"{o}del Centenary
 Meeting, Vienna, 27th-29th April 2006, phys/0612253 (2006)

\bibitem{SF} M. F. Barnsley, {\it{Superfractals}}, Cambridge
 University Press (2006) 
\bibitem{JX} J. Xia,  Ann. Math. 135, 411 (1992)


\bibitem{Go} K. G${\ddot{o}}$del, Monatsh. Math. Phys. 38, 173 (1931)

\bibitem{DC} N. C. da Costa and F. Doria, Int. J. Theor. Phys. 30, 1041 (1991)\\
 N. C. da Costa and F. Doria, Found. Phys. Letts. 4, 363 (1991)  

\bibitem{Ko} A. Komar, {\it{Undecidabilty of Macroscopically Distinguishable States in Quantum Field Theory}}, Phys. Rev. 133,n. 2b, 542 (1963

\bibitem{GH} R. Geroch and J. Hartle, {\it{Computability and Physical Theories}}, Found. of Phys. 16, 533 (1986) 
\end{thebibliography}
\end{document}